\documentclass[a4paper,10pt,twoside]{cpc-hepnp}

\usepackage{multicol}
\usepackage{graphicx}
\usepackage{booktabs}
\usepackage{amssymb,bm,mathrsfs,bbm,amscd}
\usepackage[tbtags]{amsmath}
\usepackage{lastpage}
\usepackage{caption}
\usepackage{fancyhdr}

\begin{document}

\pagestyle{fancy}
\lhead{}
\rhead{}
\chead{Submitted to ¡®Chinese Physics C}

\footnotetext[0]{Received 15 June 2014}
\tolerance=1
\emergencystretch=\maxdimen
\hyphenpenalty=10000
\hbadness=10000

\title{	An accurate measurement of PMT¡¯s TTS based on the photoelectron spectrum\thanks{Supported by Natural Science Foundation of China(11275196).  }}

\author{%
HUANG Wei-Ping
\quad TANG Ze-Bo
\quad LI Cheng$^{1)}$\email{licheng@ustc.edu.cn}
\quad JIANG Kun
\quad ZHAO XIAO-KUN%
}
\maketitle
\thispagestyle{fancy}

\address{%
State Key Laboratory of Particle Detection and Electronics, Department of Modern Physics,\\
 University of Science and Technology of China,  Hefei  230026,  China\\
}

\begin{abstract}
The water Cherenkov detector array (WCDA) for the large high altitude air shower observatory (LHAASO) will employ more than 3600 hemisphere 8¡¯ photomultiplier tubes (PMT). The good time performance of PMT, especially the transit time spread (TTS), is required for WCDA. TTS is usually defined as the TTS of single photoelectron, and usually determined by using single photoelectron counting technique. A method using the photoelectron spectrum is researched for the measurement of TTS. The method is appropriate for multi-photoelectrons and makes it possible to measure the TTS of different photoelectrons at the same time. The TTS of different photoelectrons is measured for Hamamatsu R5912 with the divider circuit designed in specifically. The TTS of single photoelectron is determined to 3.4 ns at $2.6 \times 10^{6}$ and satisfies the requirement of WCDA.
\end{abstract}



\begin{multicols}{2}

\section{Introduction}

The LHAASO project is to be built at Sichuan province of China \cite{lab1}. The main scientific goals of LHAASO are searching for galactic cosmic ray origins by extensive spectroscopy investigations of gamma ray sources above 30 TeV, all sky survey for gamma ray sources at energies higher than 300 GeV and energy spectrum and composition measurements of cosmic rays over a wide range covering knees with fixed energy scale and known fluxes for all species at the low energy end. To accomplish these tasks, the proposed project consists of four detector arrays: $1 km^{2}$ extensive air shower array (KM2A), $90000  m^{2}$ water Cherenkov detector array (WCDA), $5000  m^{2}$ shower core detector array (SCDA) and wide field of view Cherenkov telescope array (WFCTA), as shown in Fig.1 \cite{lab2}.

WCDA is a high-sensitivity gamma-ray and cosmic-ray detector. The main purpose of WCDA is to survey the northern sky for the very high energy (VHE) gamma ray sources. The WCDA consists of 4 water ponds, octagonal in shape for each, inscribed to a square with size $150 \times 150 m^{2}$. The depth of the pond is about 4.5 m. Each pond is divided into 900 cells sized $5 \times 5 m^{2}$, partitioned by black plastic curtains to prevent the penetration of light yielded in neighboring cells. A PMT resides in the bottom of each cell, looking upward to collect Cherenkov lights produced by charged particles produced in the air shower. It records the arrival time and the charge of pulses \cite{lab3}. The PMT used in WCDA is Hamamatsu R5912, whose properties are close to the requirement of WCDA.

For a PMT in WCDA, the arrival time of Cherenkov light generated by an extensive air shower (EAS) is very short, depending on the depth of shower front. The shower front of EAS is achieved according to the time reconstruction of various cell detectors. This reconstruction depends on the rise time and TTS of PMT in each cell detedtor. In consideration of reconstruction accuracy, The TTS of PMT in WCDA is demanded less than 4.0 ns.

¡¡TTS is the deviation in transit time of photoelectrons between the photocathode and the anode. The dispersion of transit time in a PMT is caused by the following sources of time uncertainties:\\
(1)Variation of the transit time of photoelectrons between a photocathode and a dynode due to a different initial velocity and emission angles of photoelectrons;\\
(2)Different transit times of photoelectrons emission from different points at the photocathode;\\
(3)Time spread in the electron multiplication \cite{lab4}.

From a Monte Carlo simulation of gammas from Crab nebula, the number of photoelectrons distribution for each PMT is shown in Fig. 2. The distribution is mainly concentrated on 1 photoelectron (43\%) , 2 photoelectrons (18\%) and 3 photoelectrons (10\%)\cite{lab5}. The detection of these photoelectrons makes it important to measure the TTS of these photoelectrons accurately.

 TTS of a PMT is usually determined by using single photoelectron counting technique calling for an exactly single photoelectron. The disadvantage of this method is to measure the TTS of single photoelectron only. To measure the TTS of various photoelectrons, an accurate method using the photoelectron spectrum is studied for the TTS measurement. The method is appropriate for multi-photoelectrons and makes it possible to measure the TTS of different photoelectrons at the same time. Then the relationship of TTS and number of photoelectrons is studied.

\begin{center}
\includegraphics[width=6cm]{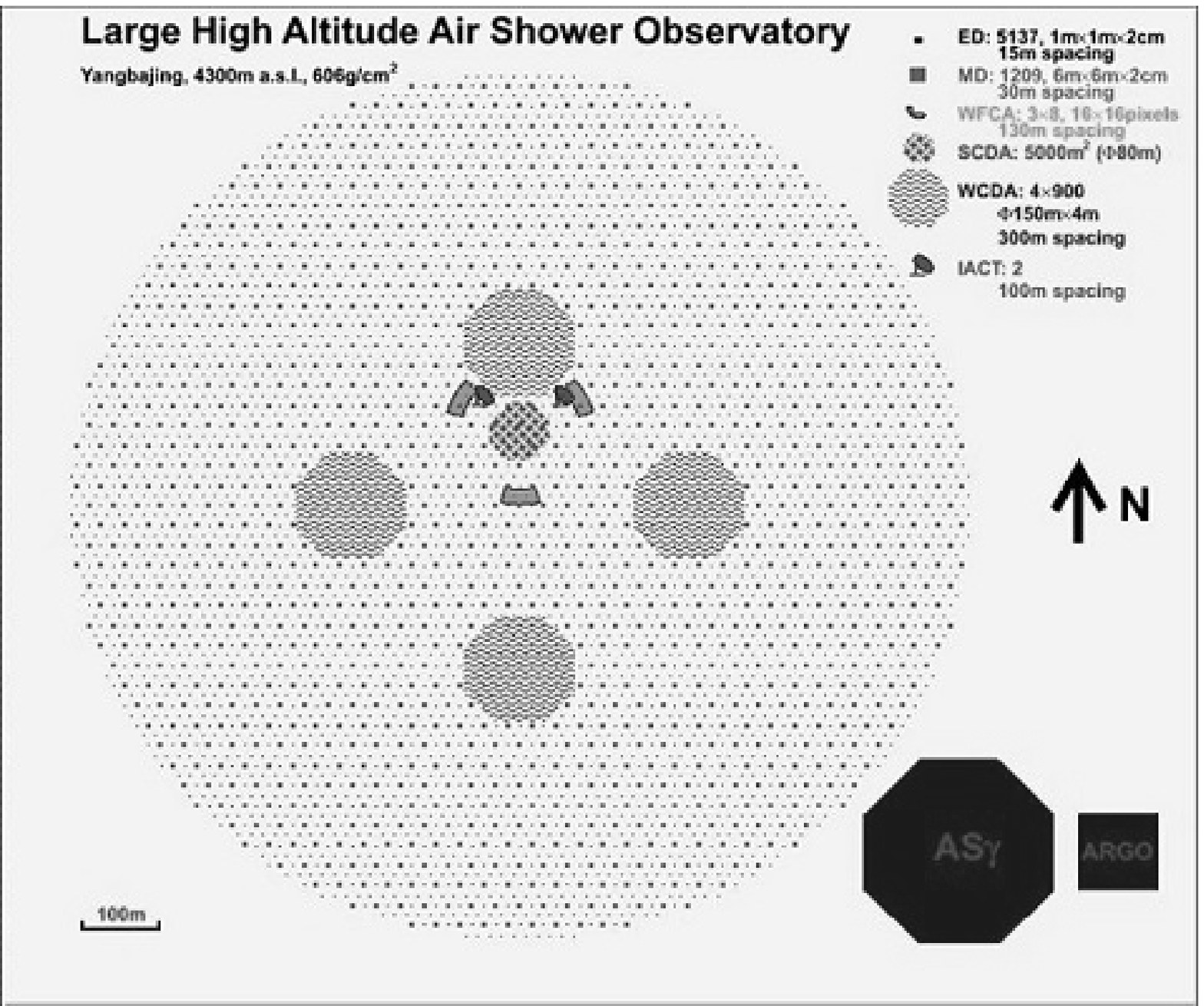}
\figcaption{\label{fig1}Layout of LHAASO.}
\end{center}

\begin{center}
\includegraphics[width=6cm]{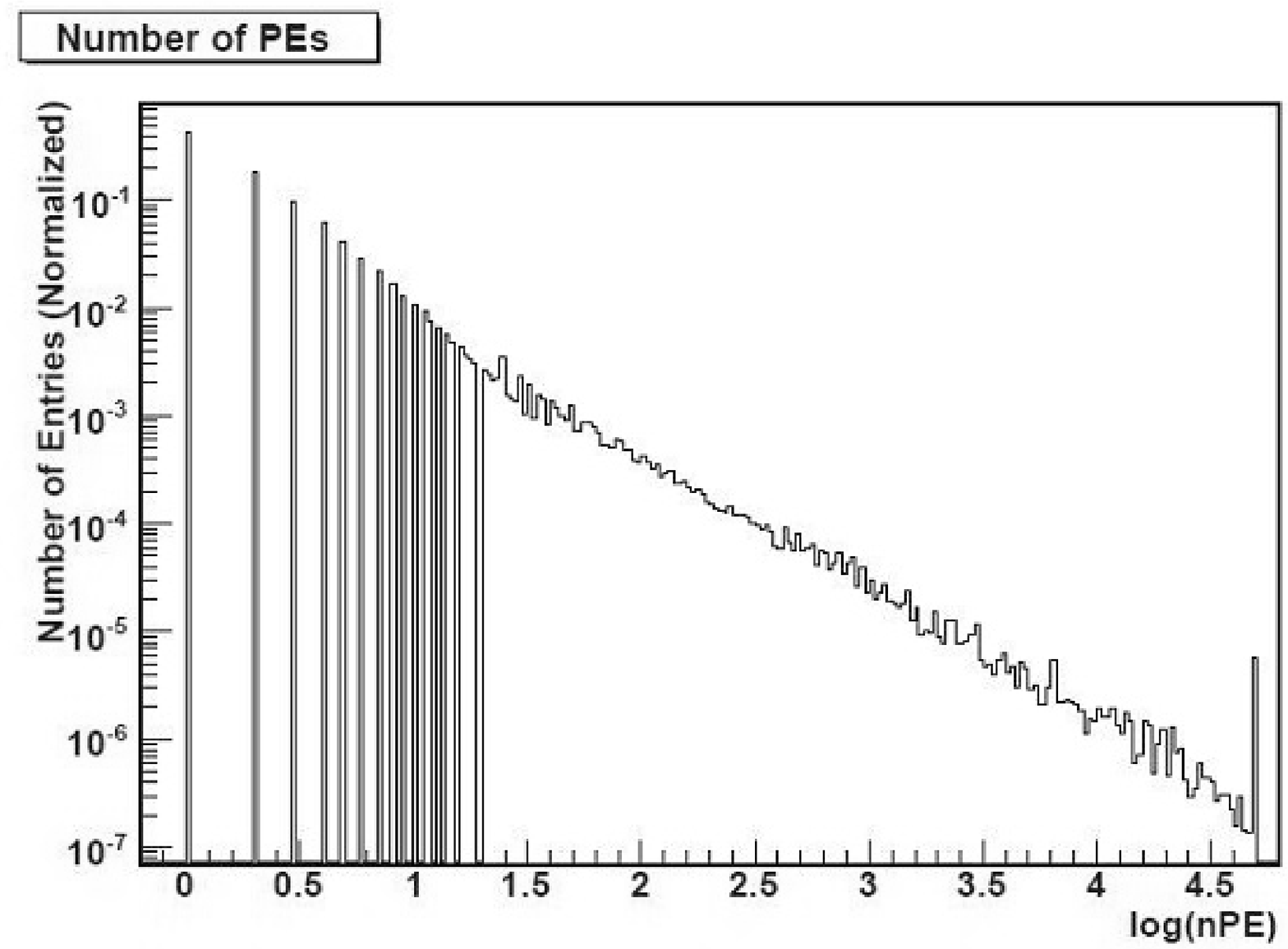}
\figcaption{\label{fig2}The nPE distribution for each PMT.}
\end{center}

\section{Experimental setup}

¡¡ A PMT base is designed for the properties measurement of R5912, whose schematic is shown in Fig. 3. The voltage distribution between the cathode, focusing electrode and first dynode is improved for the TTS performance \cite{lab6}. A typical pulse is shown in Fig. 4, in which the negative pulse is from anode and the positive from the 10th dynode. The rise time of pulse is about 4 ns and width is about 20 ns.

   An experimental setup is built for TTS measurement, shown in Fig. 5. The output of laser (Hamamatsu PLP10-040C) has a 405 nm peak and 70 ps width. The dual timer (CAEN N93B) has two sync outputs and the ADC (Lecroy 2249W) has a sensitivity of 0.25 pC. The width of ADC gate (Lecroy 2323A) is set to be 300 ns. The PMT signal from anode is split into two. One is fed into the ADC to get the charge spectrum, and the other is fed into the discriminator connected to a TDC to get the time information.

\begin{center}
\includegraphics[width=8cm]{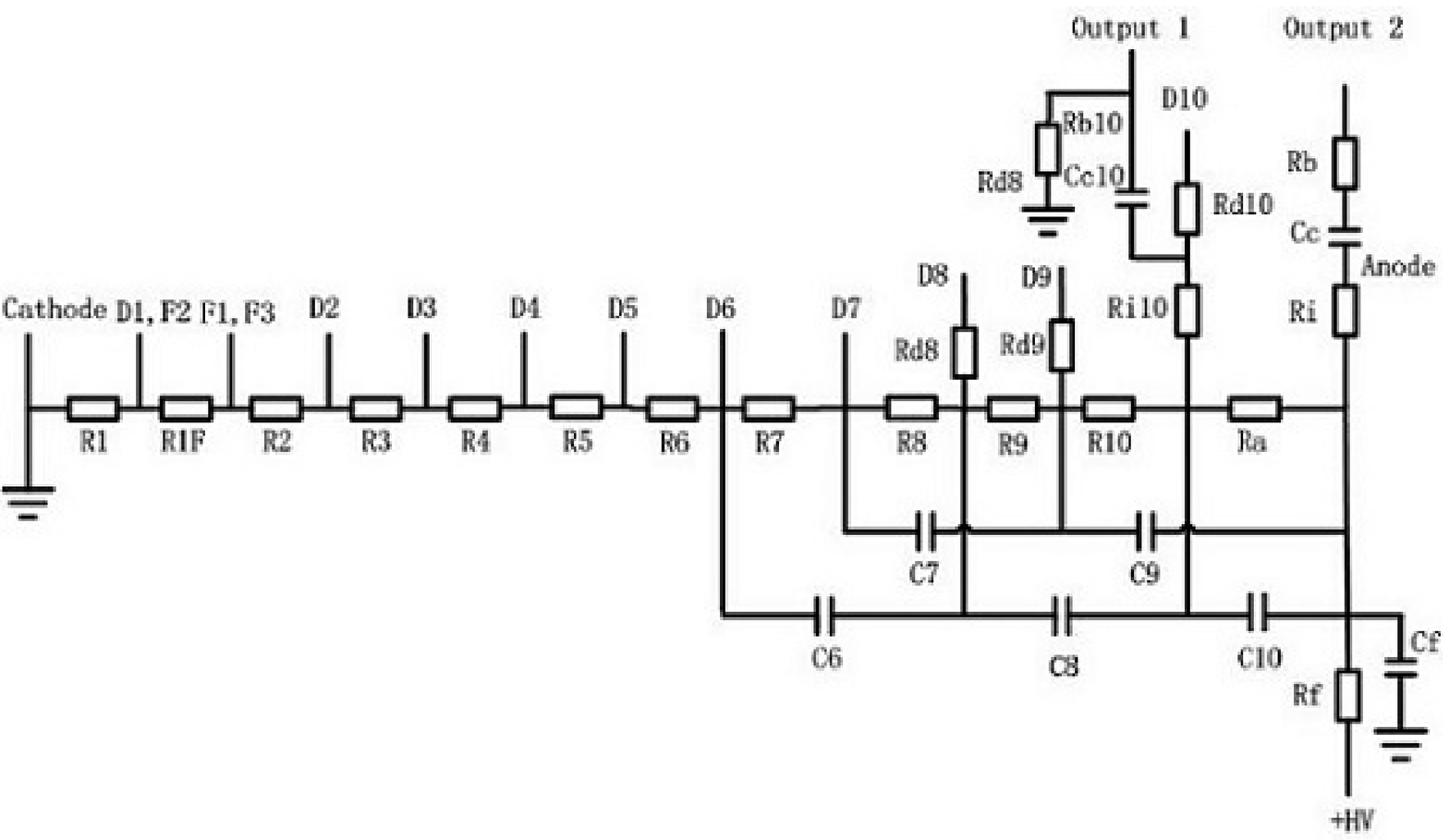}
\figcaption{\label{fig3} Schematic of R5912 base.}
\end{center}

\begin{center}
\includegraphics[width=6cm]{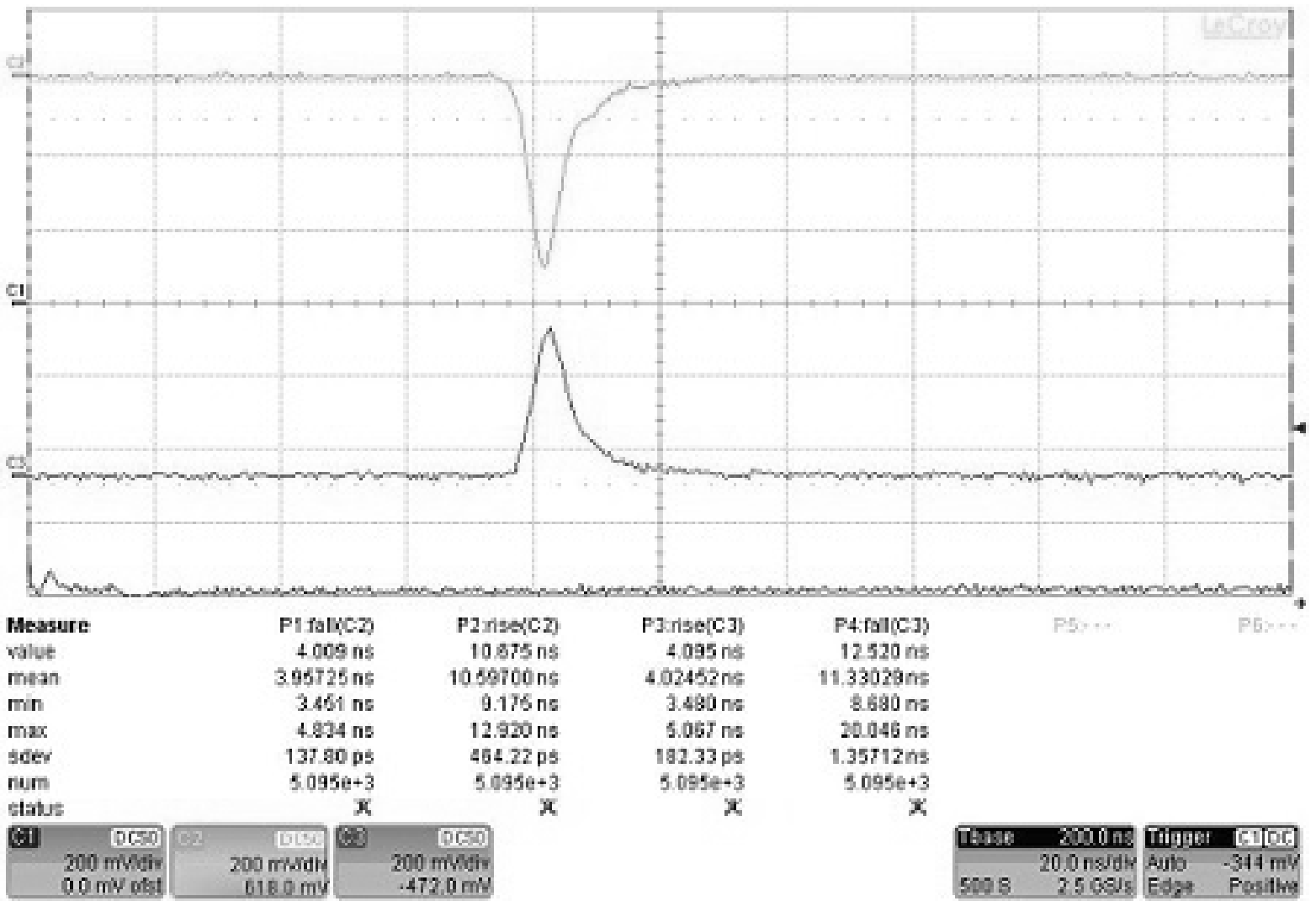}
\figcaption{\label{fig4} A typical pulse of PMT. The negative pulse is from anode and positive from 10th dynode. The number of incident photoelectron is about 290.}
\end{center}

\begin{center}
\includegraphics[width=8cm]{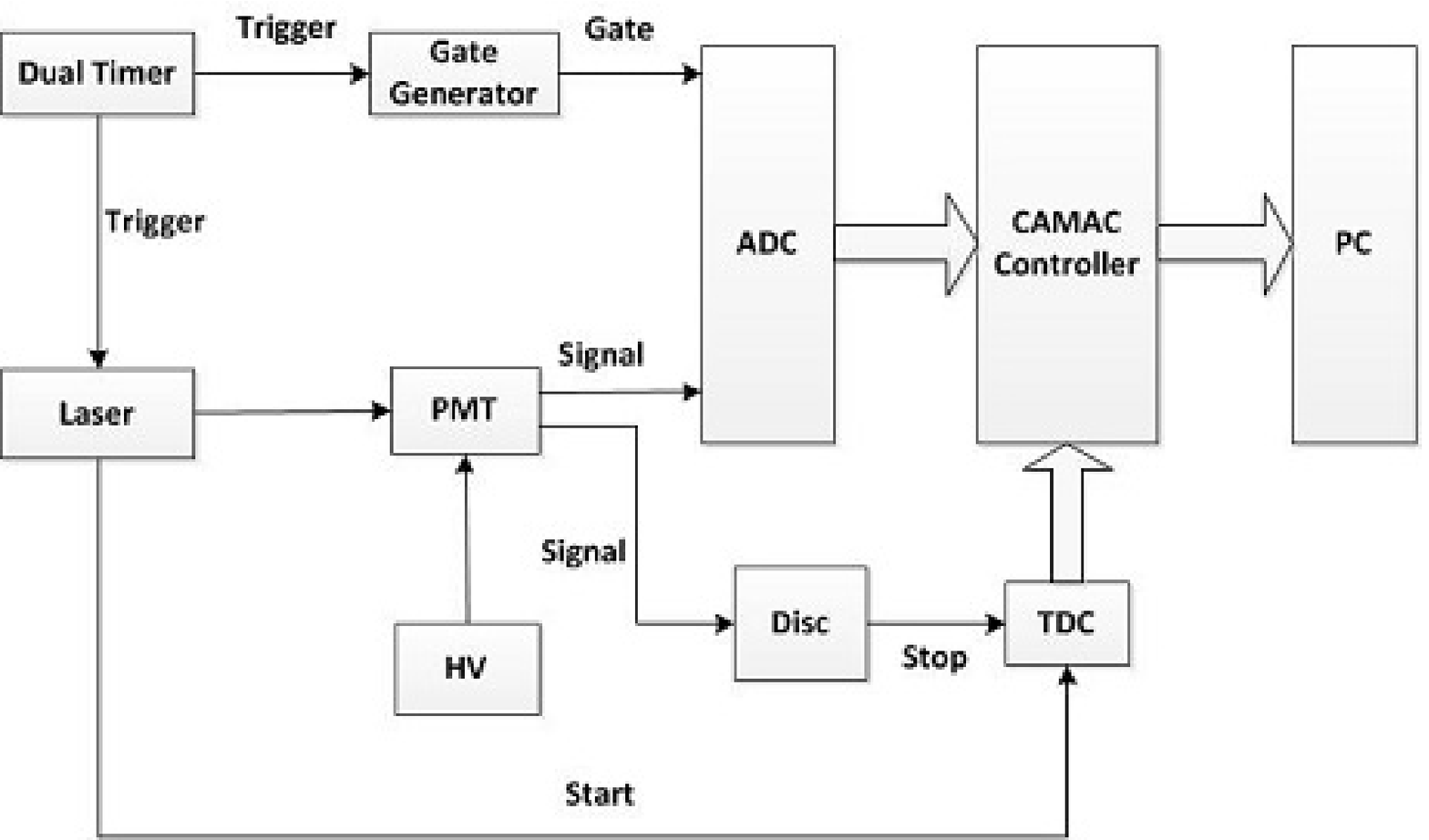}
\figcaption{\label{fig5} Experimental setup for TTS measurement.}
\end{center}

\section{TTS measurement with SPE spectrum}

The traditional method measuring TTS is the single photoelectron technique. This method needs an accurate SPE spectrum. The photoelectron spectrum is the convolution of charge distribution and pedestal, which can be described by the following distribution \cite{lab7}:

\begin{eqnarray}
 \label{eq1}
 S_{ideal}(x)&=p(0,\lambda)\times\frac{1}{\sigma_{0} \sqrt{2\pi}}e^{-\frac{(x-\mu_{0})^{2}}{2\sigma_{0}^{2}}}\\ \nonumber
 &+\sum\limits^{\infty}_{n=1}p(n,\lambda)\times\frac{1}{\sigma \sqrt{2\pi n}}e^{-\frac{(x-n\mu)^{2}}{2n\sigma^{2}}}
 \end{eqnarray}

where $\lambda$ is the mean number of photoelectrons collected by the first dynode, $ p(n;\lambda)$ is the probability, $n$ photoelectrons will be observed when their mean is $\lambda$, $x$ is the variable charge, $\mu$ is the average charge at the PMT output when one electron is collected by the first dynode and $\sigma$ is the corresponding standard deviation of the charge distribution.

In measurement, the single photoelectron spectrum can be determined by the following method: If the intensity of light reaching the PMT is such that 90 \% of the time no signal is observed, the probability of seeing no photoelectron is $ p(0,\lambda) =  e^{-\lambda} = 0.9$, so $\lambda = 0.105$. The probability of seeing a single photoelectron is $p(1,\lambda)=\lambda e^{-\lambda}=0.095$. Therefore, $p(n,\lambda) = 1 - 0.9 - 0.095 = 0.005$ when $n>1$. Consequently, any signal observed above pedstal is dominated by single photoelectron events, since $p(1,\lambda)/p(n,\lambda) = 21$. So when $\lambda < 0.105$, the spectrum is an accurate SPE spectrum. Fig. 6 shows a SPE spectrum at 1000 V with an amplifier whose amplification is 31 \cite{lab8}. The $\lambda$ is 0.023.

  Since the used discriminator is a leading edge discriminator, with the same threshold, the TDC time has a jitter because of the different charge of the PMT pulse. To correct this jitter, a time-amplitude (T-A) correction is taken, shown in Fig. 7. The relationship of TDC time and ADC charge is fit with a polynomial, then all the TDC time is corrected by this polynomial. Fig. 8 shows the distribution of transit time after T-A correction. The TTS is defined as the FWHM of this distribution.

\begin{center}
\includegraphics[width=8cm]{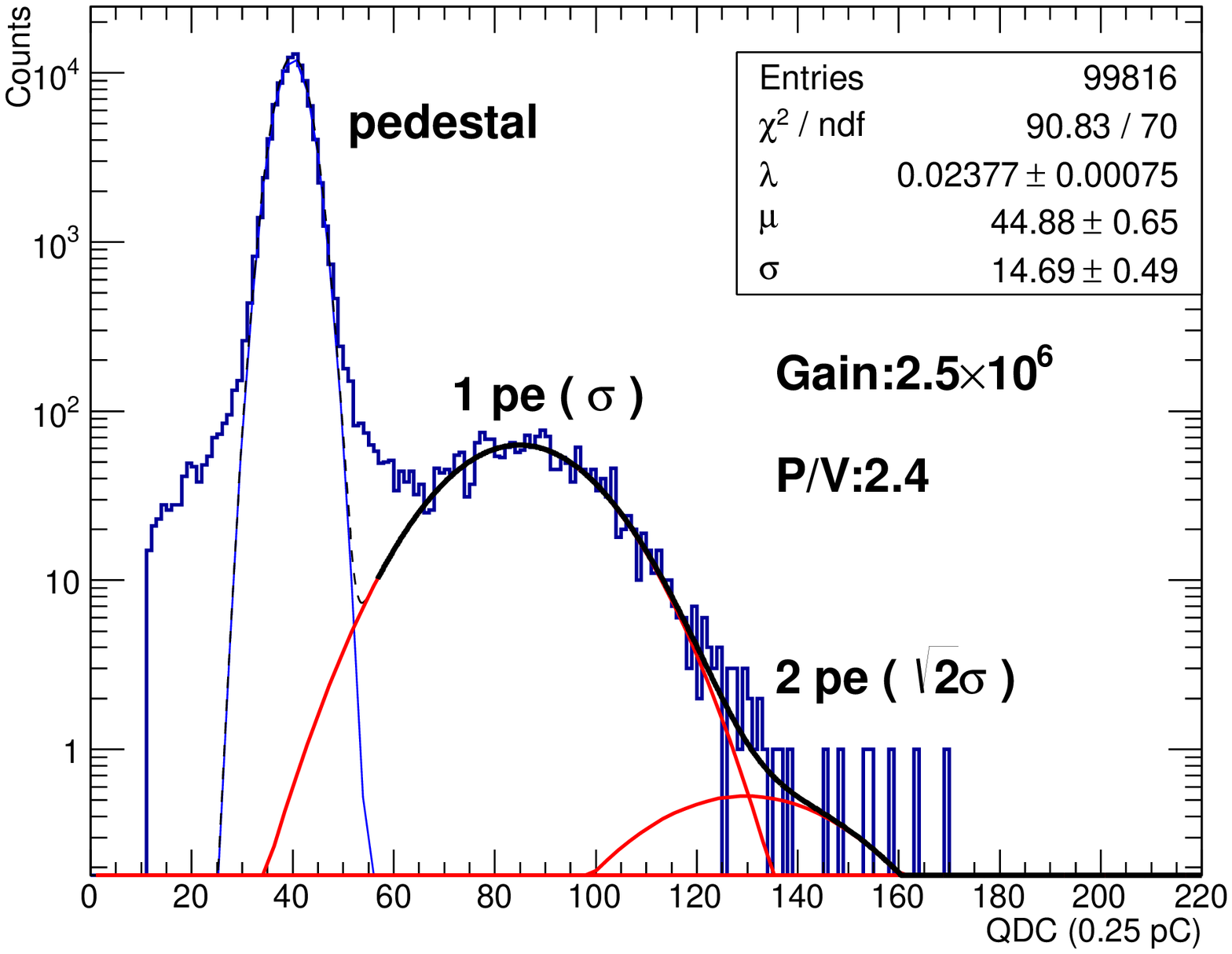}
\figcaption{\label{fig6} A typical SPE spectrum at 1000 V. The leftmost Gaussian distribution is the pedestal and the right Gaussian distributions correspond to 1, 2 photoelectrons. The P/V is the peak-valley ratio of the SPE spectrum. The gain is calculated to be $2.6\times10^{6}$. }
\end{center}

\begin{center}
\includegraphics[width=8cm]{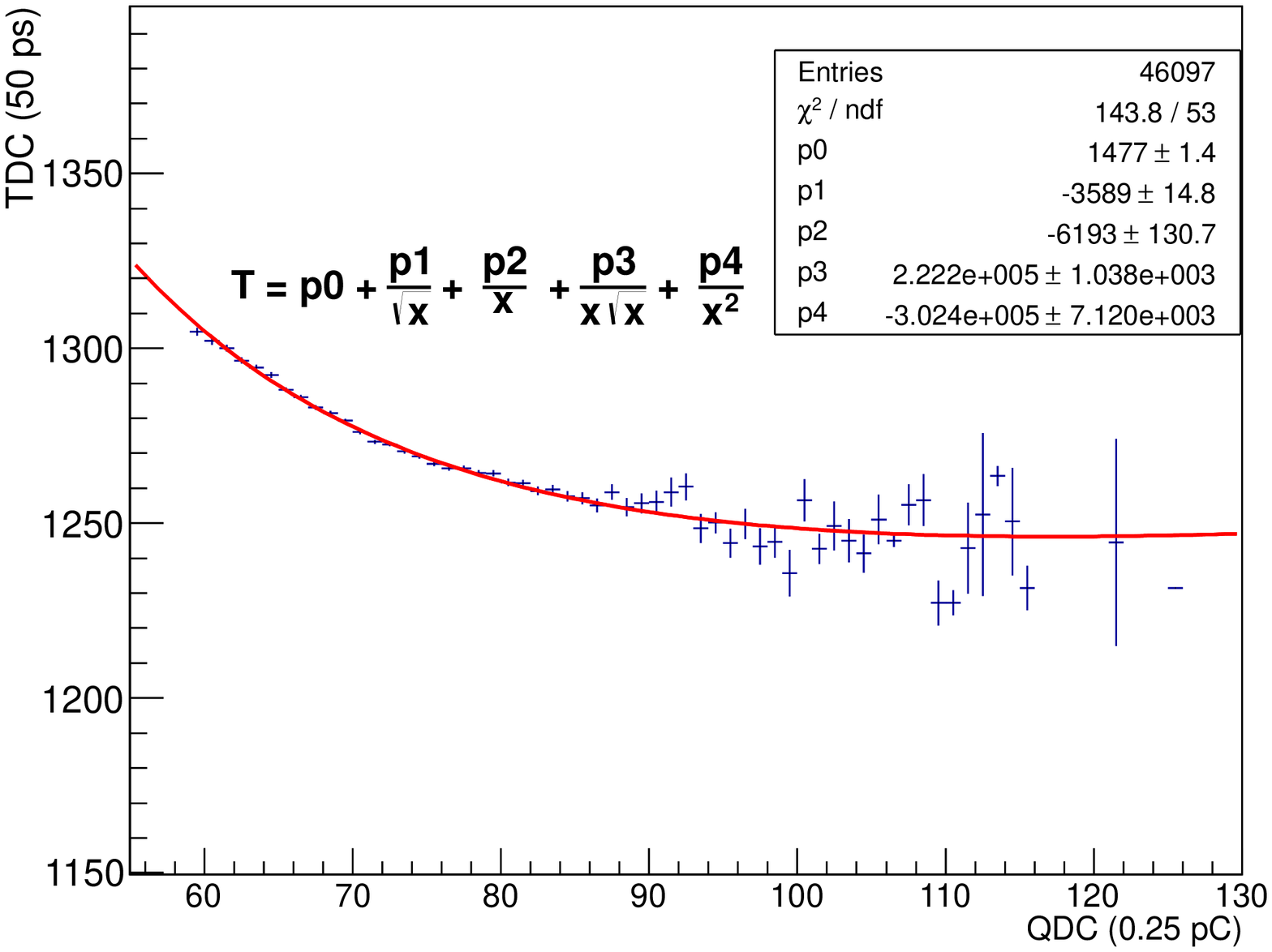}
\figcaption{\label{fig7} T-A correction.}
\end{center}

\begin{center}
\includegraphics[width=8cm]{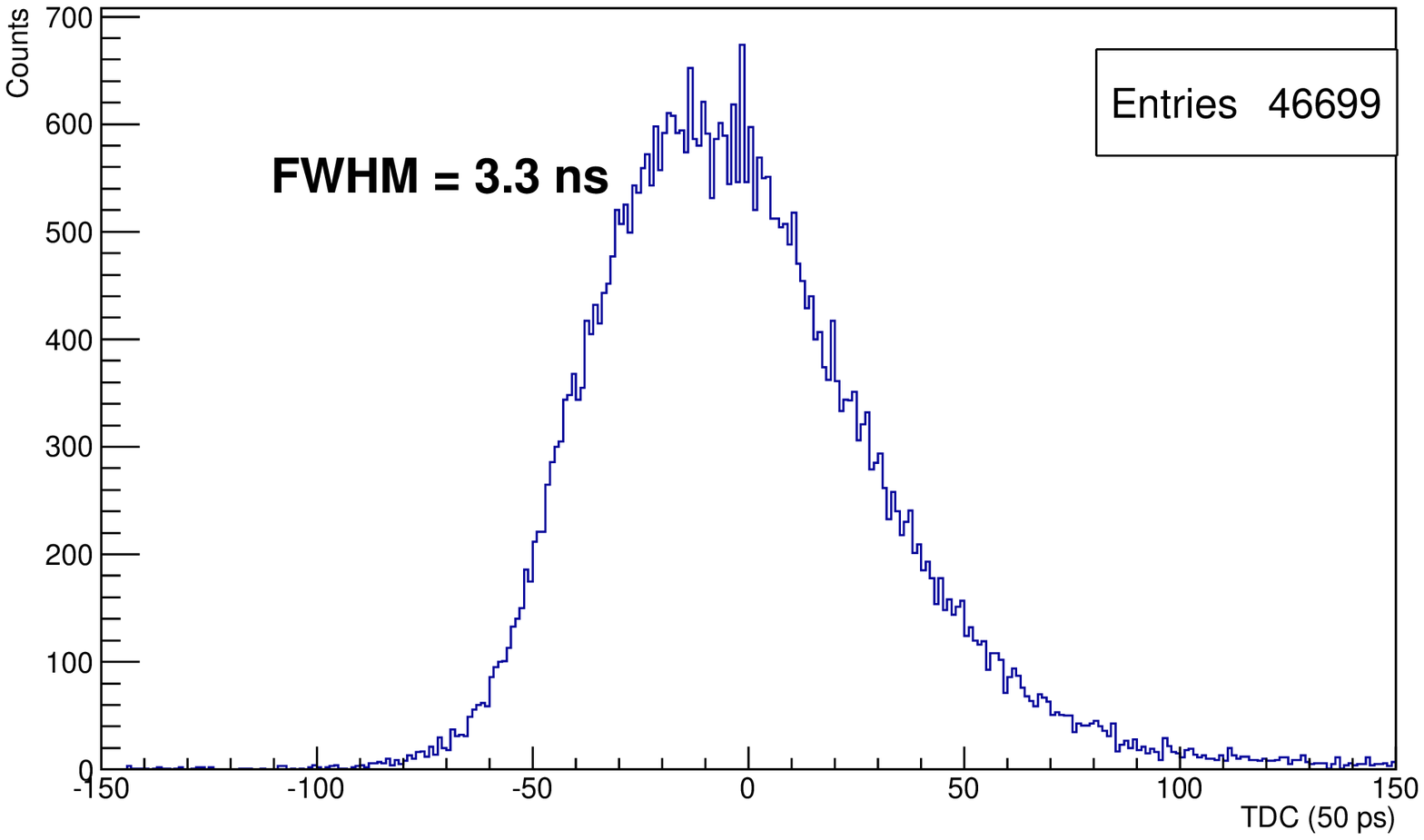}
\figcaption{\label{fig8} TTS with a real single photoelectron. The TTS is calculated to be 3.8 ns.}
\end{center}

\section{TTS measurement with multi-photoelectron spectrum}

 The single photoelectron technique is only for the TTS of single photoelectron. The number of photoelectron received by each PMT in WCDA is mainly 1,2 and 3 photoelectrons. A new method is researched to measure the TTS of different photoelectrons. The method uses the multi-photoelectron spectrum to make it possible to separate the distribution of different photoelectrons. Fig. 9 shows the divided distribution with the
 multi-photoelectron spectrum. A charge (QDC) cut is adopted to separate the single, double, triple, four photoelectrons and more. Fig. 10 shows the ratio of different photoelectrons with charge. Then the QDC region is selected to be 60$\sim$70 (Q1) for single photoelectron, 85$\sim$95 (Q2) for double photoelectrons, 115$\sim$125 (Q3) for triple photoelectrons and 140$\sim$150 (Q4) for four photoelectrons. Fig. 11 shows the distribution of transit time of different photoelectrons in the QDC cut range.
 The fraction of different photoelectrons and the TTS in different cut ranges are shown in Table 1.
 To get the real TTS of different photoelectrons, the TTS correction follows the equation:
 \begin{eqnarray}
 \label{eq2}
 {{T_{1}}^{'}}^{2}& = {A_{11}}{T_{1}}^{2} + {A_{12}}{T_{2}}^{2} \\
 {{T_{2}}^{'}}^{2}& = {A_{21}}{T_{1}}^{2} + {A_{22}}{T_{2}}^{2} + {A_{23}}{T_{3}}^{2}\\
 {{T_{3}}^{'}}^{2}& = {A_{32}}{T_{2}}^{2} + {A_{33}}{T_{3}}^{2} + {A_{34}}{T_{4}}^{2}\\
 {{T_{4}}^{'}}^{2}& = {A_{43}}{T_{3}}^{2} + {A_{44}}{T_{4}}^{2} + {A_{45}}{T_{5}}^{2}
 \end{eqnarray}
  where ${T_{n}}^{'}$ is the calculated TTS and $T_{n}$ is the real TTS of different photoelectrons. $A_{nm}$ is the normalized factor defined by the following formula:

   \begin{eqnarray}
 \label{eq3}
   A_{nm}& = {A_{nm}^{'}} / {\sum {A_{nm}^{'}} }
   \end{eqnarray}

   where $A_{nm}^{'}$ is the fraction of different photoelectrons in each cut range in Table 1, and it is assumeed that $T_{5} = 2T_{4}/\sqrt{5}$ to solve the equation. With the equation, the TTS of 1, 2, 3, 4 photoelectrons is calculated to be 3.4, 2.4, 2.0 and 1.7 ns respectively. Fig. 12 shows the relationship of TTS and the number of photoelectron and 3.4 ns is the TTS of single photoelectron. The TTS ratio of single photoelectron and $n$ photoelectrons approximates $\sqrt{n}$.

\begin{center}
\includegraphics[width=8cm]{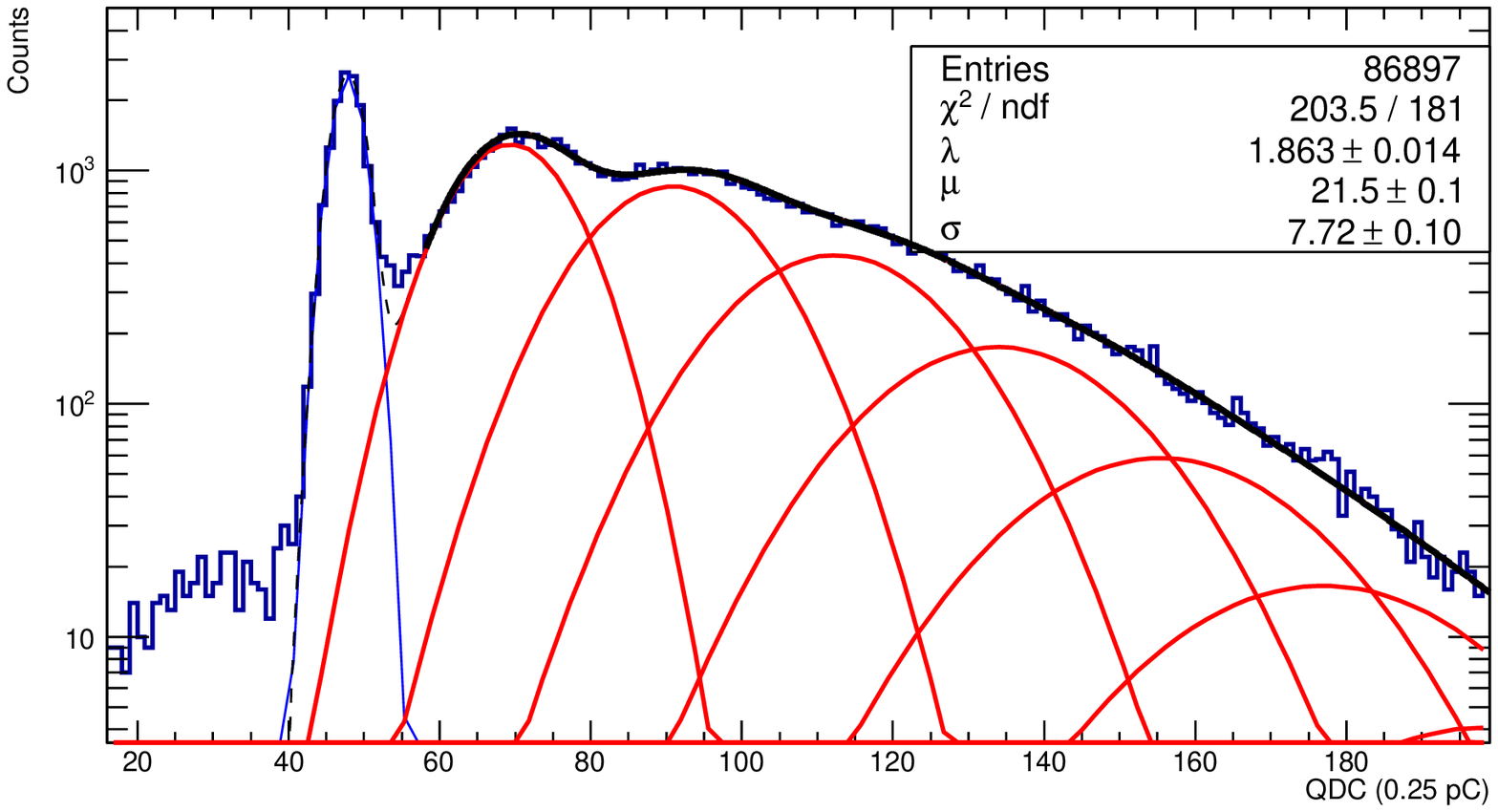}
\figcaption{\label{fig9} Multi-photoelectron spectrum in TTS measurement.}
\end{center}

\begin{center}
\includegraphics[width=8cm]{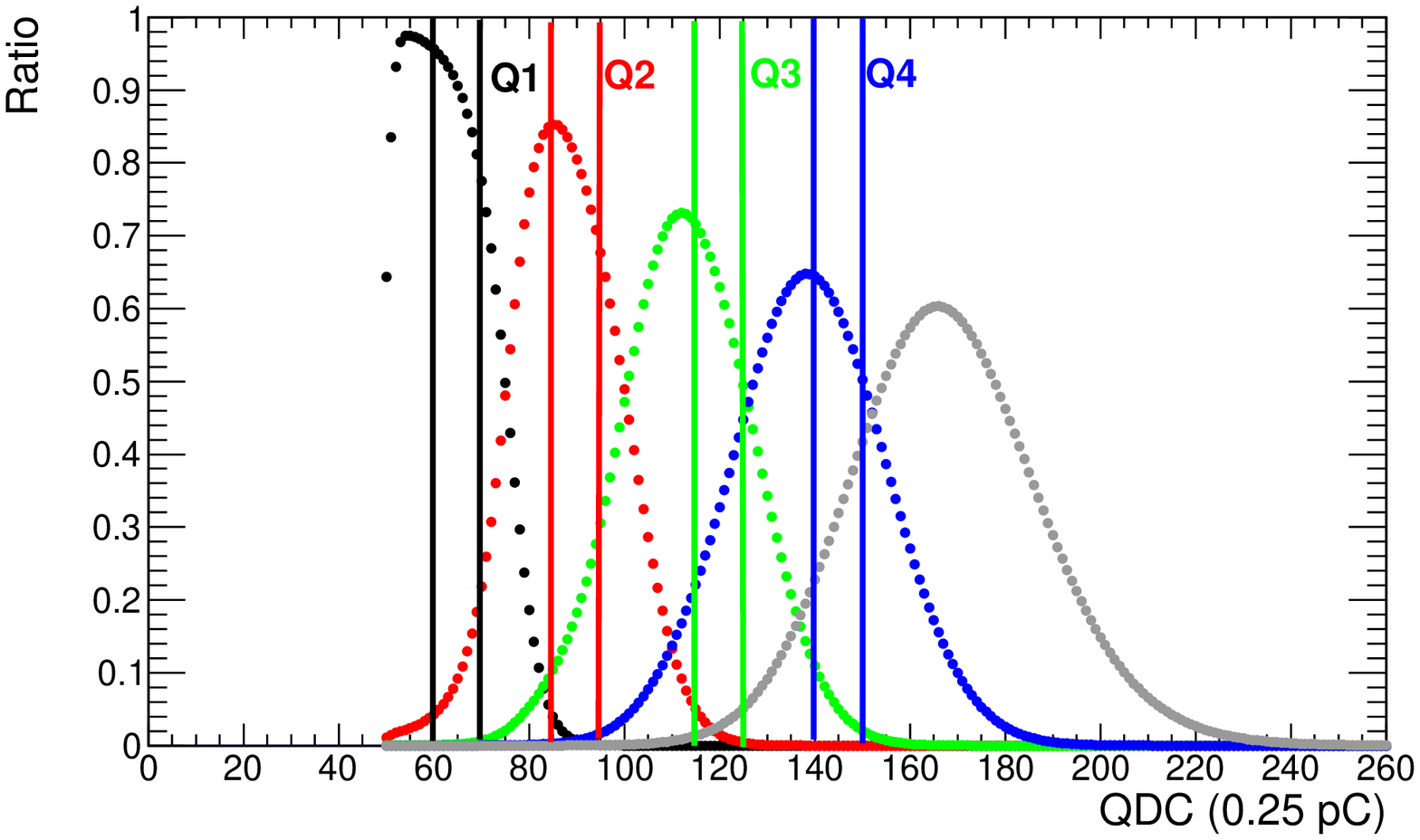}
\figcaption{\label{fig10} The ratio of different photoelectrons with charge. The single photoelectron is determined from 60 to 70 channels, the double photoelectrons from 85 to 95, triple photoelectrons from 115 to 125 and the four photoelectrons from 140 to 150.}
\end{center}

\begin{center}
\includegraphics[width=8cm]{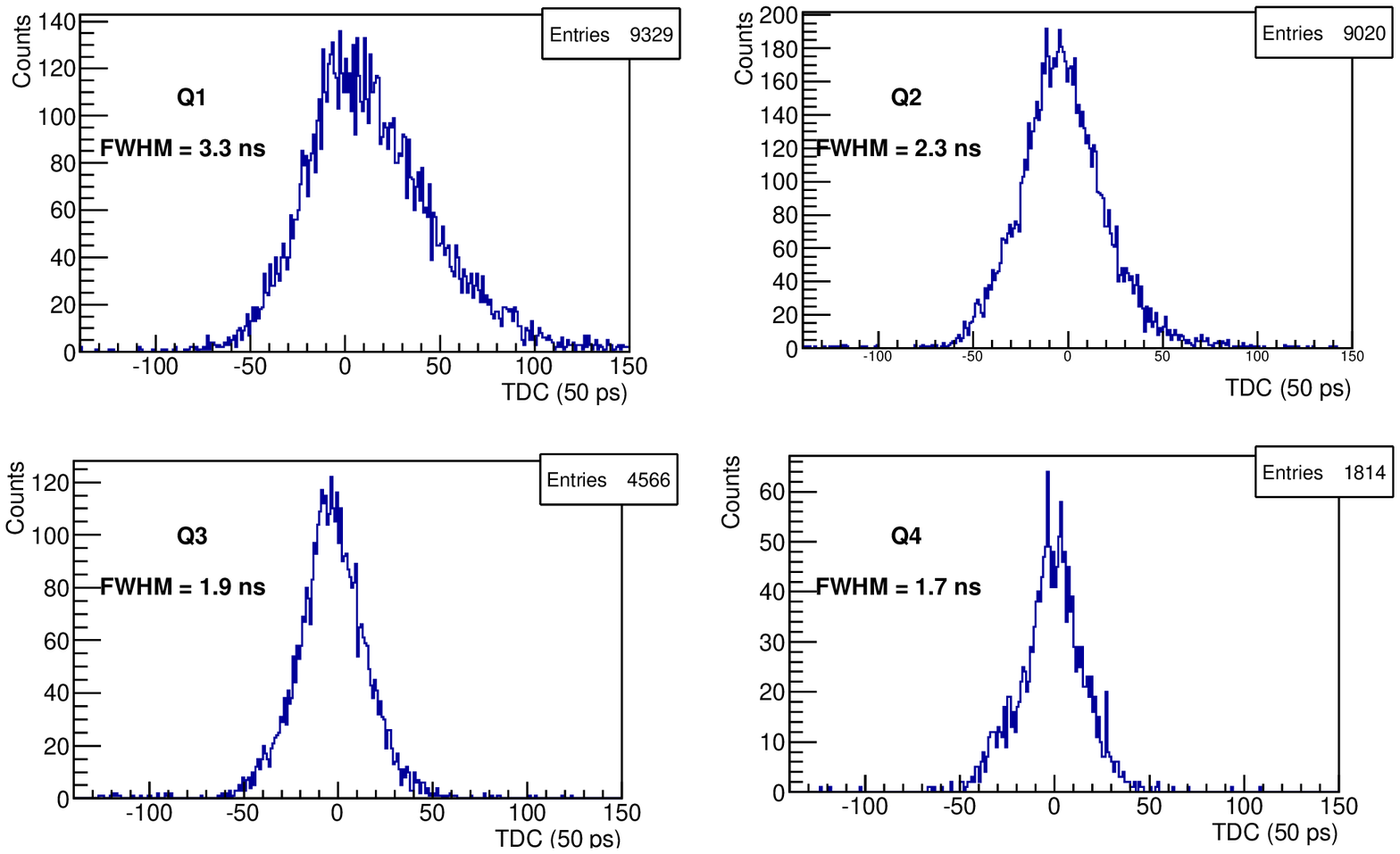}
\figcaption{\label{fig11} The transit time distribution of different QDC region.}
\end{center}

\end{multicols}

\begin{center}
\includegraphics[width=8cm]{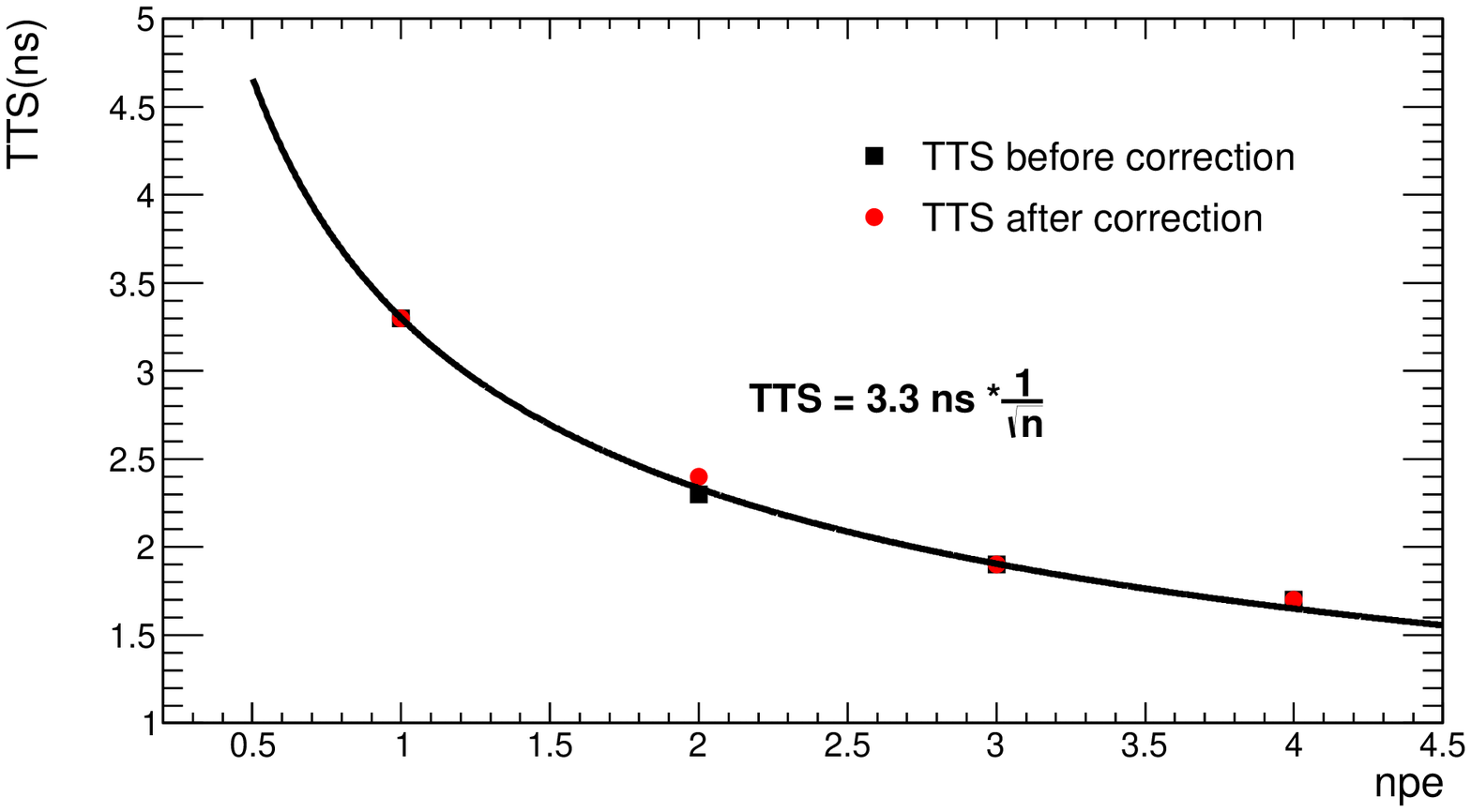}
\figcaption{\label{fig12} The relationship of TTS and number of photoelectrons.}
\end{center}

\begin{center}
\tabcaption{ \label{tab1}  The fraction of different photoelectrons and TTS in QDC cut range.}
\footnotesize
\begin{tabular*}{170mm}{@{\extracolsep{\fill}}cccc}
\toprule  QDC region & Ratio of different photoelectrons & TTS before correction/ns & TTS after correction/ns\\
\hline
60 $\sim$ 70 & 1 pe (88.9\%) + 2 pe (10.9\%) & 3.3 & 3.4 \\
85 $\sim$ 95 & 1 pe (1.0\%) + 2 pe (79.1\%) + 3 pe (18.9\%) & 2.3 & 2.4 \\
115 $\sim$ 125 & 2 pe (3.5\%) + 3 pe (62.8\%) + 4 pe (32.3\%) & 1.9 & 2.0 \\
140 $\sim$ 150 & 3 pe (6.2\%) + 4 pe (61.4\%) + 5 pe (32.4\%) & 1.7 & 1.7 \\
\bottomrule
\end{tabular*}
\end{center}

\begin{multicols}{2}

\section{Summary}
¡¡A new method based on the photoelectron spectrum to measure the TTS of different photoelectrons is studied. The advantage of the method is that the TTS of different photoelectrons is achieved at the same time. After the correction, The TTS of single, double, triple and four photoelectrons is determined to 3.4 ns, 2.4 ns, 2.0 ns and 1.7 ns, which is in inverse proportion to the square root of the number of photoelectrons. TTS of single photoelectron is consistent with the single photoelectron method. TTS of various high voltages is shown in Fig. 13, from which the measured value is less than the value provided by the manufacturer at the same voltage \cite{lab9}. This method and technique will be used to batch test of PMT performance for LHAASO WCDA. \\

\begin{center}
\includegraphics[width=8cm]{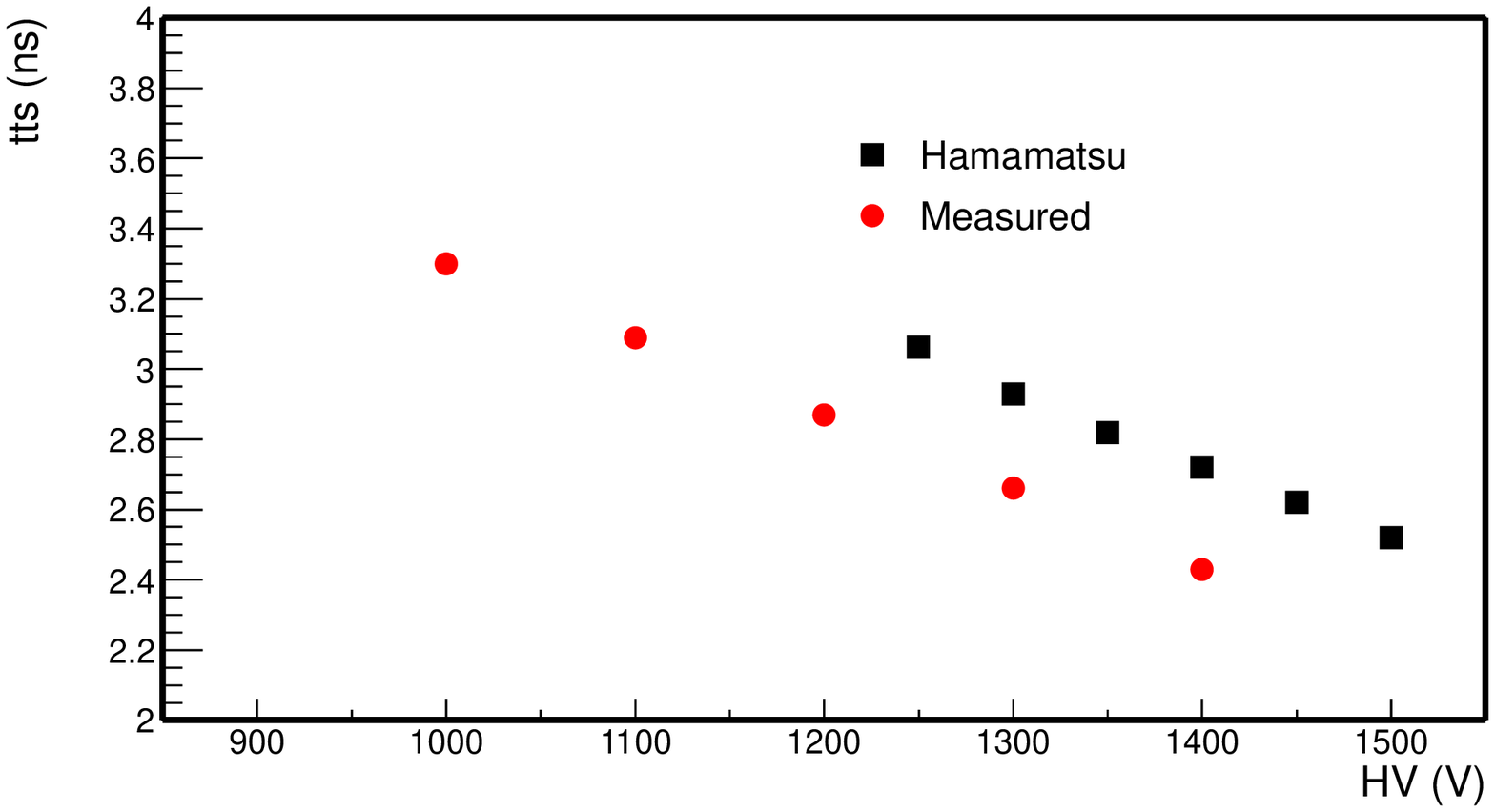}
\figcaption{\label{fig13} TTS at different voltages, compared with manufacturer.}
\end{center}

\acknowledgments{We gratefully acknowledge Prof. Zhen Cao, Prof. Huihai He, Prof. Zhiguo Yao, Dr. Mingjun Chen and other members of the LHAASO Collaboration of IHEP for their earnest support and help.}

\end{multicols}

\vspace{10mm}

\vspace{-1mm}
\centerline{\rule{80mm}{0.1pt}}
\vspace{2mm}

\begin{multicols}{2}

\end{multicols}

\clearpage


\begin{thebibliography}{90}

\vspace{3mm}

\bibitem{lab1}CAO Zhen et al. (LHAASO collaboration). Chinese Physics C, 2010, 34(2): 249.

\bibitem{lab2}Cao Zhen et al, (LHAASO collaboration). 2013, 33rd ICRC.

\bibitem{lab3}YANG Qun-Yu et al. (LHAASO collaboration). Nuclear Instruments and Methods in Physics Research A, 2011, 644: 11

\bibitem{lab4}YAO Zhi-Guo, WU Han-Rong, CHEN Ming-Jun et al. 2011, 32th ICRC.

\bibitem{lab5}Hamamatsu Corporation. Photomultiplier tubes basics and applications.

\bibitem{lab6}HUANG Wei-Ping, JIANG Kun, LI Cheng et al. Chinese Physics C, 2013, 37(3): 036001.

\bibitem{lab7}Bellamy E H, Bellettini G, Budagov J et al. Nuclear Instruments and Methods in Physics Research A, 1994, 339: 468

\bibitem{lab8}HAO Xin-Jun, LIU Shu-Bin, ZHAO Lei et al. Nuclear Electronics $\&$ Detection Technology, 2012, 32(3): 352.

\bibitem{lab9}http://sales.hamamatsu.com/en/products/electron-tube-division/detectors/photomultiplier-tubes/part-r5912.php

\end{thebibliography}
\end{document}